\documentclass[12pt,thmsa]{article}
\usepackage{sw20lart}

\input{tcilatex}
\begin{document}

\title{\textbf{Critical fluctuations, intermittent dynamics and Tsallis statistics}}
\author{A. Robledo \thanks{%
E-mail address: robledo@fisica.unam.mx } \\
Instituto de F\'{i}sica,\\
Universidad Nacional Aut\'{o}noma de M\'{e}xico,\\
Apartado Postal 20-364, M\'{e}xico 01000 D.F., Mexico.}
\date{.}
\maketitle

\begin{abstract}
It is pointed out that the dynamics of the order parameter at a thermal
critical point obeys the precepts of the nonextensive Tsallis statistics. We
arrive at this conclusion by putting together two well-defined
statistical-mechanical developments. The first is that critical fluctuations
are correctly described by the dynamics of an intermittent nonlinear map.
The second is that intermittency in the neighborhood of a tangent
bifurcation in such map rigorously obeys nonextensive statistics. We comment
on the implications of this result.


\end{abstract}

\section{Introduction}

The modern theory of critical phenomena \cite{critical1}-\cite{critical4}
has a distinguished history \cite{critical4} of achievements in
understanding the source of the scaling and universality associated to
continuous phase transitions. The introduction of renormalization group (RG)
concepts provided the means for calculating exponents, scaling functions and
marginal dimensionalities of critical points in thermal and other
statistical-mechanical models. Moreover, the RG methods have been fruitful
in other areas of physics, in condensed matter problems, in non-linear
dynamics and in other fields \cite{critical4}. The success of the RG
strategy in handling problems involving many length scales is illustrated by
the use of a coarse-grained free energy or effective action. For instance,
the equilibrium configurations of Ising spins at the critical temperature
shows magnetic domains on all size scales and these are suitably studied by
means of the Landau-Ginzburg-Wilson (LGW) continuous spin model \cite
{critical1}-\cite{critical4}.

Here we may add yet other universal aspect to critical point properties, the
nonextensivity of order-parameter fluctuations (as explained in more detail
below and in Ref. \cite{robledo1}). This previously unidentified property
develops as the size of subsystems or domains of a thermal system at its
critical point is allowed to become infinitely large. These domains have
been shown \cite{athens1}-\cite{athens4}, via the use of the LGW free
energy, to possess intermittency properties, and these, in turn, are seen to
comply \cite{robledo2} \cite{baldovin1} with the presumptions of the
nonextensive generalization \cite{tsallis0} \cite{tsallis1} of the
Boltzmann-Gibbs (BG) statistical mechanics.

In a series of papers \cite{athens1}-\cite{athens4} a bridge has been built
connecting the equilibrium dynamics of fluctuations at an ordinary thermal
critical point with the intermittent dynamics of critical nonlinear maps at
a tangent bifurcation. With the initial aim of investigating the origin of
the relationship between the fractal geometry of clusters of the order
parameter and the critical exponents of the phase transition, a theoretical
approach was devised that lead to the evaluation of the partition function
of a cluster or domain where the dominant contributions arise from a
singularity (similar to an instanton) located in the space outside it \cite
{athens1} \cite{athens2}. Subsequently \cite{athens3} \cite{athens4}, a
nonlinear map for the average order parameter was constructed whose dynamics
reproduce the averages of the thermal critical properties. This map has as a
main feature the tangent bifurcation and so time evolution is intermittent.

Recently \cite{robledo2} \cite{baldovin1}, we have shown that the known
exact \textit{static} solution of the RG equations for the tangent
bifurcation in nonlinear maps also describes the \textit{dynamics} of
iterates. The fixed-point expressions have the specific form that
corresponds to the temporal evolution of ensembles of iterates prescribed by
the nonextensive formalism. The proof rests on the derivation of the
sensitivity to initial conditions $\xi _{t}$ exclusively from RG procedures
without approximations followed by comparison with the nonextensive $\xi
_{t} $. The study of the intermittency transition has been expanded via
detailed derivation of their $q$-generalized Lyapunov coefficients $\lambda
_{q}$ and interpretation of the different types of sensitivity $\xi _{t}$ 
\cite{baldovin1}. Likewise, the properties of the intricate trajectories at
the edge of chaos in unimodal maps have been analytically obtained leading
too to the determination of $\lambda _{q}$ and interpretation of the
dynamics at the strange attractor \cite{baldovin2} \cite{mayoral1}.

We combine the results mentioned in the previous two paragraphs to point out
the nonextensive nature of critical fluctuations. This conclusion relates to
the quest of the physical circumstances for which BG statistics fails to be
applicable and its nonextensive generalization might offer correct
descriptions. These anomalous situations are signalled by the vanishing of
the Lyapunov coefficients (a single coefficient $\lambda _{1}=0$ for
one-dimensional maps) and exhibit nonergodicity or unusual phase space
mixing \cite{robledo2}-\cite{baldovin2} \cite{baldovin3}. At an
intermittency transition hindered or incomplete mixing in phase space arises
from the special 'tangency' shape of the map at its origin. This has the
effect of confining or expelling trajectories causing irregular phase-space
sampling, in contrast to the thorough coverage in generic states with $%
\lambda _{1}>0$. The occurrence of anomalous nonextensive states appears to
be related with a nonuniform convergence of limits, such as the
thermodynamic and infinite time limits. Here we comment on the nonuniform
convergence associated to the description of critical domains as obtained
from the LGW free energy.

Below we expand our arguments.

\section{Criticality and intermittency}

We briefly recall basic elements and results of the approach in Refs. \cite
{athens1}-\cite{athens4}. The starting point is the partition function of
the $d$-dimensional system at criticality, 
\begin{equation}
Z=\int D[\phi ]\exp (-\Gamma _{c}[\phi ]),  \label{partition1}
\end{equation}
where 
\begin{equation}
\Gamma _{c}[\phi ]=g_{1}\int_{\Omega }dV\left[ \frac{1}{2}(\nabla \phi
)^{2}+g_{2}\left| \phi \right| ^{\delta +1}\right]  \label{landau1}
\end{equation}
is the critical LGW free energy of a system of $d$-dimensional volume $%
\Omega $, $\phi $ is the order parameter (e.g. magnetization per unit
volume) and $\delta $ is the critical isotherm exponent. The partition
function $Z$ was evaluated for a subsystem of size $V\sim R^{d}$ and by
approximating the path integral in Eq. (\ref{partition1}) as a summation
over the saddle-point configurations of $\Gamma _{c}[\phi ]$ - an
approximation valid for $g_{1}\gg 1$. Integration of the Euler-Lagrange
equation associated to the saddle points of $\Gamma _{c}[\phi ]$, and
identification of dominant saddle points, leads to power-law magnetization
profiles for (spherically symmetric) critical clusters $\phi (r)$, and,
finally, to the evaluation of the free energy $\Gamma _{c}[\phi ]$ and the
partition function $Z$ in closed form \cite{athens1} \cite{athens2}. In
doing this it was important to notice that only configurations with $%
r_{0}\gg R$, where $r_{0}$ is a system-dependent reference position $%
r_{0}=r_{0}(g_{2},\delta ,\phi (0))$, have a nonvanishing contribution to
the path integration \cite{athens1} \cite{athens2}. These configurations
vanish for the infinite size system, so there is nonuniform convergence in
relation to the limits $R\rightarrow \infty $ and $r_{0}\rightarrow \infty $%
, a feature that is significant for our connection with $q$-statistics.
Based on the above results the fractal geometry of the critical clusters was
determined and its relationship with the exponent $\delta $ was derived. The
power-law dependence of the magnetization on the cluster radius was
identified with the fractal dimension of the geometry of the cluster \cite
{athens1} \cite{athens2}. Multifractal properties appear when global, rather
than a single cluster, properties are considered.

Subsequent to this development, a link was revealed \cite{athens3} \cite
{athens4} between the fluctuation properties of a critical system described
by Eq. (\ref{partition1}) and the dynamics of marginally chaotic
intermittent maps. By considering the space-averaged magnetization $\Phi
=\int_{V}\phi (x)dV$, the statistical weight 
\begin{equation}
\rho (\Phi )=\exp (-\Gamma _{c}[\Phi ])/Z,  \label{invariant1}
\end{equation}
where $\Gamma _{c}[\Phi ]\sim g_{1}g_{2}\Phi ^{\delta +1}$ and $Z=\int d\Phi
\exp (-\Gamma _{c}[\Phi ])$, was seen to be the invariant density of a
statistically equivalent one-dimensional map. The functional form of this
map was obtained as the solution of an inverse Frobenius-Perron problem \cite
{athens3} \cite{athens5}. For small values of $\Phi $ the map has the form 
\begin{equation}
\Phi _{n+1}=\Phi _{n}+u\Phi _{n}^{\delta +1}+\epsilon ,  \label{tangent1}
\end{equation}
where the amplitude $u$ depends on $g_{1}$, $g_{2}$and $\delta $, and the
shift parameter $\epsilon \sim R^{-d}$. Eq. (\ref{tangent1}) can be
recognized as that describing the intermittency route to chaos in the
vicinity of a tangent bifurcation \cite{schuster1}. The complete form of the
map displays a superexponentially decreasing region that takes back the
iterate close to the origin in one step. Thus the parameters of the thermal
system determine the dynamics of the map. Averages made of order-parameter
critical configurations are equivalent to iteration time averages along the
trajectories of the map close to the tangent bifurcation. The mean number of
iterations in the laminar region was seen to be related to the mean
magnetization within a critical cluster of radius $R$. There is a
corresponding power law dependence of the duration of the laminar region on
the shift parameter $\epsilon $ of the map \cite{athens3}. For $\epsilon >0$
the (small) Lyapunov exponent is simply related to the critical exponent $%
\delta $ \cite{athens2}.

\section{ Intermittency and nonextensivity}

Next, in a few words, we recall the nonextensive nature of the nonlinear
dynamics at the tangent bifurcation \cite{robledo2} \cite{baldovin1}. At
this transition, the intermittency route to chaos, the ordinary Lyapunov
exponent $\lambda _{1}$ vanishes and the sensitivity to initial conditions $%
\xi _{t}\equiv \left| dx_{t}/dx_{0}\right| $ (where $x_{t}$ is the orbit
position at time $t$ given the initial position $x_{0}$ at time $t=0$) is no
longer given by the BG law $\xi _{t}=\exp (\lambda _{1}t)$ but acquires
either a power or a super-exponential law \cite{baldovin1}. The nonextensive
formalism indicates that $\xi _{t}$ is given by the $q$-exponential
expression, 
\begin{equation}
\xi _{t}=\exp _{q}(\lambda _{q}t)\equiv [1-(q-1)\lambda _{q}t]^{-1/(q-1)},
\label{qsensitivity}
\end{equation}
containing the entropic index $q$ and the $q$-generalized Lyapunov
coefficient $\lambda _{q}$. Also, according to the generalized theory, the $%
q $-Pesin identity $K_{q}=\lambda _{q}$ replaces the ordinary Pesin identity 
$K_{1}=\lambda _{1}$, $\lambda _{1}>0$, where $K_{q}\equiv t^{-1}\left[
S_{q}(0)-S_{q}(t)\right] $ and $K_{1}\equiv t^{-1}[S_{1}(t)-S_{1}(0)]$ are
the rates of entropy increment based on the Tsallis entropy 
\begin{equation}
S_{q}\equiv \sum_{i}p_{i}\ln _{q}\left( \frac{1}{p_{i}}\right) =\frac{%
1-\sum_{i}^{W}p_{i}^{q}}{q-1},  \label{tsallis}
\end{equation}
(where $\ln _{q}y\equiv (y^{1-q}-1)/(1-q)$ is the inverse of $\exp _{q}(y)$)
and the BG entropy 
\begin{equation}
S_{1}(t)=-\sum_{i=1}^{W}p_{i}(t)\ln p_{i}(t),  \label{BG1}
\end{equation}
respectively. (Above, $p_{i}(t)$ is the distribution obtained from the
relative frequencies with which the positions of an ensemble of trajectories
occur within cells $i=1,...,W$ at iteration time $t$). In the limit $%
q\rightarrow 1$ the expressions for the nonextensive theory reduce to the
ordinary BG expressions. See Ref. \cite{baldovin1} for a more rigorous
description of the Pesin identity and related issues.

Assisted by the known RG treatment for the tangent bifurcation \cite
{schuster1}, the formula for $\xi _{t}$ has been rigorously derived \cite
{robledo2} \cite{baldovin1} and found to comply with Eq. (\ref{qsensitivity}%
). Also the validity of the $q$-Pesin identity has been substantiated for
this problem \cite{robledo2}. The tangent bifurcation is usually studied by
means of the map 
\begin{equation}
f(x)=\epsilon +x+u\left| x\right| ^{z}+O(\left| x\right| ^{z}),\ u>0,
\label{n-thf1}
\end{equation}
in the limit $\epsilon \rightarrow 0$. The associated RG fixed-point map $%
x^{\prime }=f^{*}(x)$ was found to be 
\begin{equation}
x^{\prime }=x\exp _{z}(ux^{z-1})=x[1-(z-1)ux^{z-1}]^{-1/(z-1)},\epsilon =0,
\label{fixed1}
\end{equation}
as it satisfies $f^{*}(f^{*}(x))=\alpha ^{-1}f^{*}(\alpha x)$ with $\alpha
=2^{1/(z-1)}$ and has a power-series expansion in $x$ that coincides with
Eq. (\ref{n-thf1}) in the two lowest-order terms. (Above $x^{z-1}\equiv
\left| x\right| ^{z-1}\mathrm{sgn}(x)$). The long time dynamics is readily
derived from the static solution Eq. (\ref{fixed1}), one obtains 
\begin{equation}
\xi _{t}(x_{0})=[1-(z-1)ax_{0}^{z-1}t]^{-z/(z-1)},  \label{sensitivity2}
\end{equation}
and so, $q=2-z^{-1}$ and $\lambda _{q}(x_{0})=zax_{0}^{z-1}$ \cite{robledo2} 
\cite{baldovin1}. When $q>1$ the left-hand side ($x<0$) of the tangent
bifurcation map Eq. (\ref{n-thf1}) exhibits a weak insensitivity to initial
conditions, i.e. power-law convergence of orbits. However at the right-hand
side ($x>0$) of the bifurcation the argument of the $q$-exponential becomes
positive and this results in a `super-strong' sensitivity to initial
conditions, i.e. a sensitivity that is faster than exponential \cite
{baldovin1}.

\section{Nonextensivity and criticality}

The implications of joining the results described in the previous two
sections are apparent. In the subsystem of infinite size $R\rightarrow
\infty $ the dynamics of critical fluctuations obey the nonextensive
statistics. This is expressed via the time series of the average order
parameter $\Phi _{n}$, i.e. trajectories $\Phi _{n}$ with close initial
values separate in a superexponential fashion according to Eq. (\ref
{sensitivity2}) with $q=(2\delta +1)/(\delta +1)>1$ and with a $q$-Lyapunov
coefficient $\lambda _{q}$ determined by the system parameter values $\delta 
$, $g_{1}$, $g_{2}$ and $\Phi _{0}$ \cite{robledo1}. Also, when considering
an ensemble of trajectories $\{\Phi _{n}\}$ with prescribed distribution of
initial conditions, the $q$-Pesin identity $K_{q}=\lambda _{q}$ holds with
the rate of entropy production $K_{q}$ evaluated according to the
nonextensive entropy $S_{q}$.

It is interesting to comment on the conditions for the incidence of $q$%
-statistical properties at criticality and the manner in which these
develop. The order parameter profile for a large but finite-size domain $%
R>>1 $ has the form \cite{athens1}-\cite{athens3} 
\begin{equation}
\phi (r)=A_{d}(r^{2}-r_{0}^{2})^{(2-d)/2},d\geq 3,
\end{equation}
that, parenthetically, can be rewritten in terms of a $q$-exponential. There
is a singularity in $\phi (r)$ when $r=r_{0}$, the reference position. We
keep in mind the requirement $R\ll r_{0}$ for the critical clusters $\phi
(r) $ to be of relevance to the partition function $Z$ and also the map
shift parameter dependence on the domain size $\epsilon \sim R^{-d}$. The
time evolution of $\Phi $ displays laminar episodes of duration $<n>\sim
\epsilon ^{-\delta /(\delta +1)}$ and the Lyapunov coefficient in this
regime is $\lambda _{1}\sim \epsilon $ \cite{athens2}. Within the first
laminar episode the dynamical evolution of $\Phi $ obeys $q$-statistics, but
for very large times the occurrence of many different laminar episodes leads
to an increasingly chaotic orbit consistent with the small $\lambda _{1}>0$
and BG statistics is recovered. As $R$ increases ($R\ll r_{0}$ always) the
time duration of the nonextensive regime increases and in the limit $%
R\rightarrow \infty $ there is only one infinitely long laminar nonextensive
episode with $\lambda _{1}=0$ and with no crossover to BG statistics. On the
other hand when $R>r_{0}$ the clusters $\phi (r)$ are no longer dominant,
for the infinite subsystem $R\rightarrow \infty $ their contribution to $Z$
vanishes \cite{athens1}-\cite{athens3} and no departure from BG statistics
is expected to occur.

This study is developed in Ref. \cite{robledo1}.

\section{Current questions on $q$-statistics}

What is the significance of the connections we have presented? Are there
other connections between critical phenomena and transitions to chaos? Are
all critical states - infinite correlation length with vanishing Lyapunov
coefficients - outside BG statistics? When does BG statistics stop working?
Where, and in that case why, does Tsallis statistics apply? Is ergodicity
failure the basic playground for applicability of generalized statistics? Is
there a link between critical dynamics and glassy dynamics?

Suggestive results and partial answers to these questions are given here and
in the cited references. In relation to this it is pertinent to mention the
following recent development. In Ref. \cite{robledo3} it is argued that the
dynamics at the noise-perturbed edge of chaos in logistic maps is analogous
to that observed in supercooled liquids close to vitrification. That is, the
three major features of glassy dynamics in structural glass formers,
two-step relaxation, aging, and a relationship between relaxation time and
configurational entropy, are displayed by orbits with $\lambda _{1}=0$. Time
evolution is obtained from the Feigenbaum RG transformation, it is expressed
analytically via $q$-exponentials, and described in terms of nonextensive
statistics.\medskip

\textbf{Acknowledgments.} It is a great pleasure to dedicate this work to
Constantino Tsallis on the occasion of his 60th birthday. I am grateful for
hospitality offered at Angra dos Reis, Brazil, and for partial support by
DGAPA-UNAM and CONACyT (Mexican Agencies).


\begin{thebibliography}{99}
\bibitem{critical1}  K.G. Wilson, Physica 73, 119 (1974).

\bibitem{critical2}  M.E. Fisher, Rev. Mod. Phys. 46, 597 (1974).

\bibitem{critical3}  K.G. Wilson, Rev. Mod. Phys. 55, 583 (1983).

\bibitem{critical4}  M.E. Fisher, Rev. Mod. Phys. 70, 653 (1998).

\bibitem{robledo1}  A. Robledo, to be submitted.

\bibitem{athens1}  N.G. Antoniou, Y.F. Contoyiannis, F.K. Diakonos, and C.G.
Papadoupoulos, Phys. Rev. Lett. 81, 4289 (1998).

\bibitem{athens2}  Y.F. Contoyiannis and F.K. Diakonos, Phys. Lett. A268,
286 (2000).

\bibitem{athens3}  N.G. Antoniou, Y.F. Contoyiannis, and F.K. Diakonos,
Phys. Rev. E 62, 3125 (2000).

\bibitem{athens4}  Y.F. Contoyiannis, F.K. Diakonos, and A. Malakis, Phys.
Rev. Lett. 89, 035701 (2002).

\bibitem{tsallis0}  C. Tsallis, J. Stat. Phys. 52, 479 (1988).

\bibitem{tsallis1}  For a recent review see, \textit{Nonextensive Entropy --
Interdisciplinary Applications}, M. Gell-Mann and C. Tsallis, eds., (Oxford
University Press, New York, 2004), in press. See
http://tsallis.cat.cbpf.br/biblio.htm for full bibliography.

\bibitem{robledo2}  A. Robledo, Physica A 314, 437 (2002); A. Robledo,
Physica D (2004, in press) and cond-mat/ 0202095.

\bibitem{baldovin1}  F. Baldovin and A. Robledo, Europhys. Lett. 60, 518
(2002).

\bibitem{baldovin2}  F. Baldovin and A. Robledo, Phys. Rev. E 66, 045104(R)
(2002); F. Baldovin and A. Robledo, cond-mat/0304410.

\bibitem{mayoral1}  E. Mayoral and A. Robledo, Physica A, submitted, and
cond-mat/0401128.

\bibitem{baldovin3}  F. Baldovin, L.G. Moyano, A.P. Majtey, A. Robledo and
C. Tsallis, Physica A, submitted, and cond-mat/0312407.

\bibitem{athens5}  F.K. Diakonos, D. Pingel, P. Schmelcher, Phys. Lett. A
264, 162 (1999).

\bibitem{schuster1}  See, for example, H.G. Schuster, \textit{Deterministic
Chaos. An Introduction}, 2nd Revised Edition (VCH Publishers, Weinheim,
1988).

\bibitem{robledo3}  A. Robledo, Physica A, submitted, and cond-mat/0307285.
\end{thebibliography}
\end{document}